\documentclass[aps,pra,twocolumn,groupedaddress]{revtex4}

\usepackage{graphicx}

\begin{document}

\title{Optimized U-MOT for experiments with ultracold atoms near
surfaces}

\author{S. Wildermuth}
\author{P. Kr\"uger}
\email[Electronic mail: ]{krueger@physi.uni-heidelberg.de}
\author{C. Becker}
\author{M. Brajdic}
\author{S. Haupt}
\author{A. Kasper}
\author{R. Folman}
\altaffiliation[Current address: ]{Department of Physics, Ben
Gurion University, Beer-Sheva 84105, Israel}
\author{J. Schmiedmayer}
\affiliation{Physikalisches Institut, Universit\"at Heidelberg,
D-69120 Heidelberg, Germany} \homepage{www.atomchip.org}

\date{\today}

\begin{abstract}
We present an integrated wire-based magnetooptical trap for the
simplified trapping and cooling of large numbers of neutral atoms
near material surfaces. With a modified U-shaped current-carrying
Cu structure we collect $>3\times 10^8$ $^{87}$Rb atoms in a
mirror MOT without using quadrupole coils. These atoms are
subsequently loaded to a Z-wire trap where they are evaporatively
cooled to a Bose-Einstein condensate close to the surface.
\end{abstract}

\pacs{39.25.+k}

\maketitle

Recent progress in the field of trapping and manipulating atoms in
micropotentials has significantly improved the possibilities of
investigating the interaction of trapped atoms with material
objects. Newly available techniques allow, for example, to bring
cold neutral atoms close to surfaces and use them as highly
sensitive local probes of electric and magnetic surface
potentials. In this context, both the effects of thermally induced
currents (Johnson noise) \cite{Hen03} and of disorder in surface
structures \cite{Wan03} have been studied theoretically. The first
indications of such disorder potentials have already been observed
experimentally \cite{Kra02,Jon03} and some effects attributed to
Johnson noise have been measured in various materials
\cite{Har03}.

Similarly, structuring surfaces allows to tailor potentials for
the atoms with a resolution of the order of the atom--surface
distance. This distance can be in the micron range, possibly
below. In this way, integrated devices for the controlled
manipulation of matter waves, so called atom chips, can be built.
Atom chips combine the potential of microfabrication technology,
i.e. to create nearly arbitrary structures for detailed and robust
atom manipulation, with the ability of a controlled quantum
evolution developed in atomic physics and quantum optics. They
pave the path to many applications ranging from fundamental
physics of mesoscopic atomic systems and issues of low
dimensionality to implementations of quantum information
processing \cite{Fol02,Rei02}.

The starting point of many of these experiments is a cloud of
ultracold atoms, ideally a Bose-Einstein condensate (BEC), close
to the surface. The cloud has to be formed in situ or be
transported to the experimental region. Here, we present an
important simplification of this process that is particularly well
suited for an efficient production of BEC samples near surfaces.
We demonstrate that a large number ($>3\times 10^8$) of $^{87}$Rb
atoms can be collected in a modified wire based magnetooptical
trap (MOT) located just millimeters above the reflecting surface
in a simple dispenser loaded setup under ultrahigh vacuum (UHV)
conditions ($<10^{-11}$mbar). This allows to subsequently transfer
the atoms into a Z-shaped wire trap \cite{Cas99,Rei99,Haa01} that
is used to cool atomic samples into the Bose-condensed phase
\cite{Sch03}.

A MOT requires laser light forces from all directions.
Consequently, placing a MOT close to a surface implies that either
the surface and the laser beam diameters have to be small enough
relative to the height of the MOT above the surface or that the
surface is transparent or reflecting. The problem of a material
object (partially) obstructing the access of the six beams used in
a conventional MOT has also been circumvented by producing the MOT
\cite{Ott01} or even the condensate \cite{Lea02} elsewhere and
transfer it to the chip by means of dynamic magnetic fields
\cite{Ott01} or optical tweezers \cite{Lea02}. The alternative is
to directly load a {\em mirror MOT} \cite{Lee96,Pfa97,Rei99,Fol00}
only millimeters away from a reflecting surface that acts as a
mirror. In this configuration, at least one of the MOT beams is
reflected off the mirror. In the simple version used in many atom
chip experiments \cite{Rei99,Fol00,Jon03}, two of the regular six
MOT beams are replaced by reflections of two beams impinging upon
the mirror \footnote{In our experiments with atom chips, the
microstructures creating the potentials manipulating the atoms are
directly fabricated into the reflecting gold surface used as the
mirror.} at an angle of 45$^\circ$. To ensure the correct
quadrupole field orientation with respect to the helicities of the
light beam pairs, the quadrupole field axis has to coincide with
one of the 45$^\circ$ light beams. Up to now, this had to be
considered a drawback since the coils usually employed to provide
the field are bulky, dissipate a large amount of power, and
deteriorate the optical access to the MOT itself and to the region
where the experiments are carried out. As experimental setups are
likely to grow more complex in the future, including quadrupole
coils in the setup will present a major obstacle. Apparatus
involving cryostats aiming at a significant reduction of thermal
current noise in conducting surfaces \cite{Hen03} are just one
example.

\begin{figure}
    \includegraphics[width = \columnwidth]{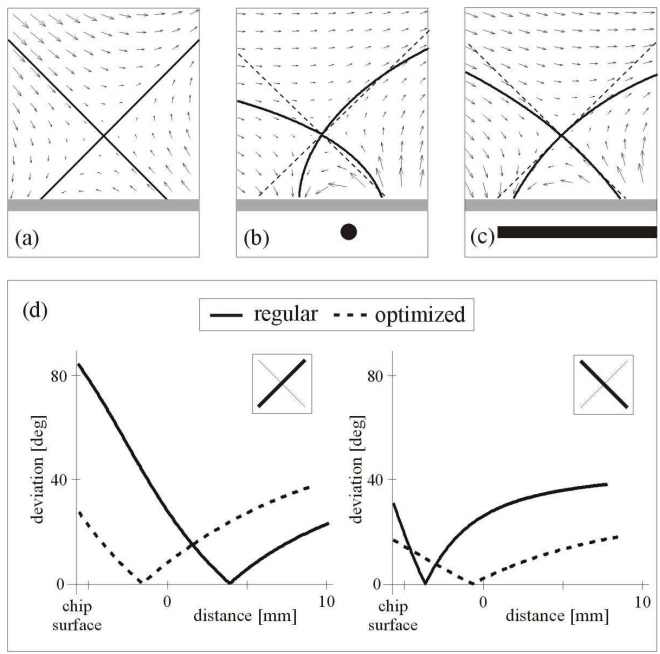}
    \caption{Vector plots of different field configurations. The
    solid (dashed) lines indicate the axes of the approximated
    (ideal) quadrupole fields. (a) ideal quadrupole field,
    (b) regular U-wire quadrupole field, (c) optimized U-wire
    quadrupole field. The wire cross section (black) and the
    surface above it (gray) are shown. (d) Angular deviations
    from the ideal quadrupole axes are plotted as a function of
    the distance from the reflecting surface along the two
    $45^\circ$ light beam paths (dashed lines in (b) and (c)). The
    solid (dashed) lines correspond to the regular (optimized) U-wire
    configuration. The zero point of the position axes is chosen to be
    the center of the quadrupole field (field zero). The broad wire U
    clearly approximates the ideal quadrupole field better throughout
    a larger spatial region than the thin wire U. The parameters
    chosen in these examples were $I_{U-wire}=55$A,
    $B_\parallel=14.5$G ($12.8$G) in the plane parallel to the wire
    and $B_\perp=0$G ($3.0$G) perpendicular to the wire for the
    regular (optimized) U. \label{fig:fields}}
\end{figure}

A known way of approximating a quadrupole field is to use a
current carrying wire that is bent in a U-shape together with a
homogenous bias field parallel to the wire plane and perpendicular
to the central bar of the U \cite{Fol02}. Fig.~\ref{fig:fields}b
shows the field configuration obtained in comparison to a field
created by external coils in the common anti-Helmholtz
configuration (Fig.~\ref{fig:fields}a). But a MOT based on a
simple U-shaped wire cannot be used for an efficient collection of
a large number of atoms (for example from the background
Rb-vapor). This is caused by the fact that the U-wire field is
only a true quadrupole field near the field center (point of
vanishing field). Further out there is a a non-vanishing angle
between the quadrupole axes and the field lines
(Fig.~\ref{fig:fields}d). This angle increases at larger distances
from the field zero, i.e. the MOT center, and eventually the
direction of the field vectors is even reversed. As the operation
principle of a MOT relies on the correct orientation of the fields
with respect to the polarization of the laser light in each beam,
the effective capture region of the trap and thus the loading rate
and the maximum number of atoms in the MOT are limited.
Consequently, the U-MOT \footnote{This simple type of U-MOT is
typically used as an intermediate experimental stage because it
can be aligned to surface patterns by construction and it allows a
simple compression of the atomic cloud as it is lowered towards
the surface by increasing the homogenous bias field.} has to be
loaded from a regular quadrupole coil MOT in order to collect a
large number of atoms.

However, by altering the geometry of the U-shaped wire, a much
better approximation of a quadrupole field can be obtained: The
bent field lines in the case of a simple U-wire can be attributed
to the fact that a thin wire produces a field whose field lines
are circles. Consequently, the simplest way to overcome this is to
fan out the current flow through the central part of the U by
replacing the thin wire by a broadened plate. Inclining the bias
field with respect to the plane formed by the outer leads of the U
improves the field configuration further. If the plate is inclined
and if the shape of the current flow through the plate is adjusted
properly, the resulting field will approximate an ideal quadrupole
field even more closely.

We chose to set the last two possibilities aside in our
experiment, mainly because they lead to only marginal improvement
compared to the wide U, and they are more difficult to implement.
Fig.~\ref{fig:fields}c shows the field vectors of the quadrupole
field obtained with a modified planar U-shaped wire. The various
parameters (geometry, wire current, and bias field) were optimized
numerically to achieve typical field gradients (10--20G/cm) of a
MOT at a height of 6--8mm above the the wire center (4--6mm above
the chip surface) while maintaining small angular deviations of
the field from an ideal quadrupole field throughout the maximal
capture region given by a typical light beam diameter of 2cm. A
comparison of the field configurations of the U-wire quadrupole
field with the ideal field shows no significant differences in the
planes not shown in Fig~\ref{fig:fields}. Only the field gradients
deviate from those obtained in a conventional quadrupole
configuration: In the direction parallel to the central bar of the
U-shaped wires, the gradients are weak while those in the
transverse directions are of approximately equal magnitude. The
gradient ratios for the regular (optimized) U-wire are $\sim
1:4:5$ ($\sim 1:3:4$), for the ideal quadrupole field $1:1:2$.
Gradient ratios, however, are not critical for a MOT operation. In
fact, this can even be an advantage because the aspect ratio of
the MOT cloud is better matched to the magnetic microtraps. With
our configuration, it turns out that moderate wire currents of
50--70A at small power consumptions ($<1$W) and small bias fields
of 7--13G are sufficient to create a near to ideal quadrupole
field at a variable height above the chip surface. The residual
angles of the field vectors are small enough to lie within the
tolerance of a MOT, as was tested by rotating the light
polarizations in an external MOT experiment. The MOT remained
unimpaired for elliptical polarizations corresponding to
deviations of the field line direction of up to $40^\circ$ from
the ideal situation.

\begin{figure}
    \includegraphics[width= \columnwidth]{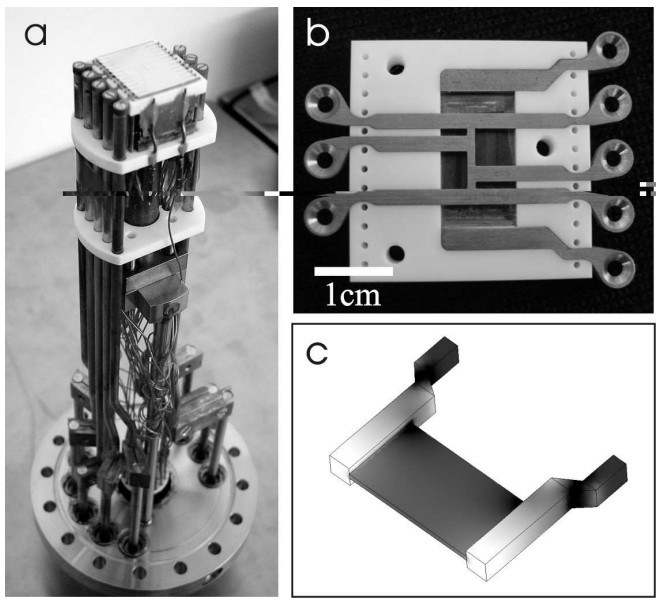}
    \caption{(a) Atom chip assembly: The U-wire structure for the MOT
    and an additional structure containing Z-shaped wires in several sizes are connected
    to high-current vacuum feedthroughs. The atom chip is mounted directly on top of these wire
    structures. (b) Photograph of the wire structures fitted into a
    MACOR ceramics holder. (c) Results of a numerical calculation
    of the current density distribution in the U-wire. Dark (light)
    shades correspond to high (low) current densities. The thick
    connecting leads ensure a homogenous fanning out of the current
    through the central plate as it is needed to improve the
    quadrupole field for the MOT.
    \label{fig:design}}
\end{figure}

In our experimental implementation we use a U-wire structure that
has been machined out of a single copper piece. This Cu structure
is incorporated in a MACOR ceramics block holding an atom chip.
Between chip and central part of the U, a small space was left to
allow the placement of another copper structure that contains
several Z-shaped wires for magnetic trapping for BEC production
(Fig.~\ref{fig:design}b). In order to keep ohmic heat dissipation
as low as possible while allowing currents of up to 100A, a wire
cross section of at least 7mm$^2$ is maintained all over the
U-wire structure. The 3mm $\times$ 3mm leads are thicker than the
plate (thickness $\times$ width $\times$ length = 0.7mm $\times$
10mm $\times$ 18mm) to ensure a homogenous current density in the
plate (Fig.~\ref{fig:design}c). Isolated by a thin (100$\mu$m)
Kapton foil, the 1mm thick additional structure for purely
magnetic trapping is positioned on top of the plate. The geometry
of this structure resembles an H with two extra leads connected to
the central bar. This allows to run currents through a variety of
Z-shaped wires with a (center to center) length of the central bar
ranging from 4mm to 10mm by choosing the proper connectors. The U-
and the H-shaped structures were designed in such a way that their
surfaces lie in a common plane so that an atom chip can be mounted
directly on top of both structures. The copper structures are
connected to high current vacuum feedthroughs by simple screw
contacts, the chip wires are attached to pin connectors by a
bonding technique.

\begin{figure}
    \includegraphics[width = 1.0 \columnwidth]{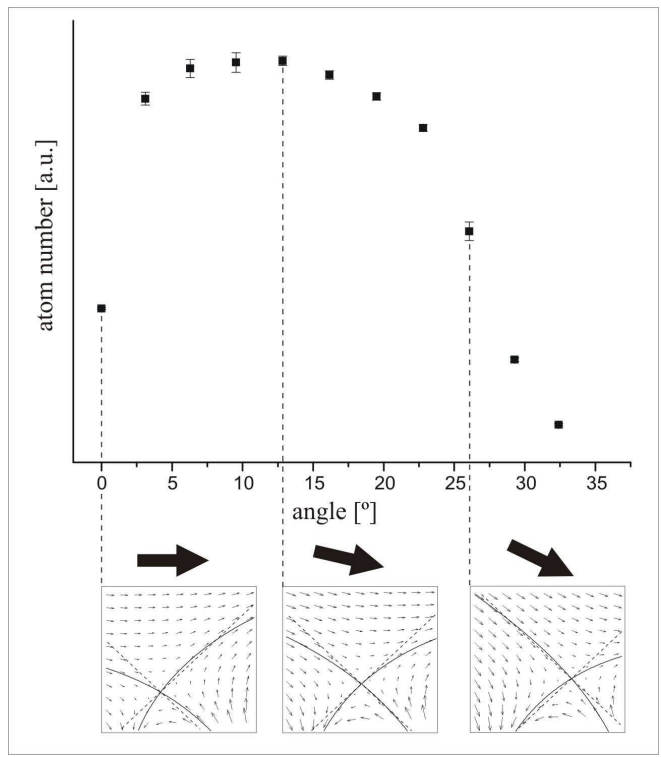}
    \caption{Top: The number of atoms is plotted versus the
tilting angle between the bias field and the plane of the U-shaped
wire. For these measurements the U-current was 55A, the bias field
strength 13G. Bottom: Corresponding vector field plots for three
different angles ($0^\circ$, $13^\circ$, $26^\circ$) as indicated
by the arrows. The maximum number of atoms is trapped at a bias
field angle of 13$^\circ$ where the shape of the field is closest
to an ideal quadrupole field.
    \label{fig:exp}}
\end{figure}

The complete assembly with the current connections, the wire
structures for the MOT and magnetic traps, and the atom chip
(Fig.~\ref{fig:design}a) is built into a UHV chamber \footnote{The
vacuum system reaches a base pressure of below $7\times
10^{-12}$mbar  and is pumped by a combination of a Ti sublimation
pump and a 300l/s ion pump.} that was constructed to allow good
optical access to the experimental region directly above the
surface of the chip. This was realized by including optical
quality quartz windows in an octogonally shaped stainless steel
body. The distance from the outer surfaces to the experimental
region is 4cm and 10cm for the directions perpendicular and
parallel to the chip surface, respectively. As a source for
rubidium atoms we use three dispensers that are connected in
parallel. A high pumping speed in combination with a pulsed
operation mode of the dispensers facilitates sufficient loading
rates of the MOT of typically $3\times 10^7$ atoms/s while the
rubidium background vapor is quickly reduced in the purely
magnetic trapping phase of the experiment.

Typically, we operate the MOT at a U-wire current of 60A and a
bias field of 13G, i.e. at magnetic field gradients of 5G/cm
(20G/cm) along the axis of weakest (strongest) confinement. In
order to confirm the effect of the improved quadrupole field on
the MOT, we have measured the number of atoms in the MOT as a
function of the angle between the bias field and the plane of the
modified U-shaped wire. Fig.~\ref{fig:exp} shows the result and
the corresponding quadrupole fields for specific angles. The
dependence of the number of atoms on the quality of the
approximation of a true wide range quadrupole field is clearly
visible: the MOT contained the highest number of atoms ($3\times
10^8$) for the optimal quadrupole field that is obtained at a
13$^\circ$ inclination of the bias field. To compare the results
of the U-MOT, we carried out test experiments with a conventional
6-beam MOT before introducing the atom chip assembly in the
apparatus. Neither the loading rates nor the maximum number of
trapped atoms exceeded those measured with the U-MOT under similar
UHV conditions. Thus we conclude that a modified U-MOT can replace
a conventional MOT completely.

After loading the U-MOT from the Rb background pressure, we turn
off the dispensers while leaving both light and magnetic fields of
the MOT on for 5 seconds. During this period, the dispensers are
cooled efficiently through their Cu rod connectors (6mm diameter)
while the pressure in the chamber is quickly reduced by the pumps.
In the next step, the atoms are molasses cooled to $\sim 30\mu$K
and optically pumped to the $|F=2,m_F=2\rangle$ state. This allows
to transfer up to $2\times 10^8$ atoms to a magnetic trap
(lifetime $>30$s) that is formed by the H-shaped integrated
Cu-structure. In the configuration used, the current runs through
the innermost possible Z-shaped path where the length of the
central bar of the Z has a length of 4mm. This trap is operated
with a current of 60A through the wire and a bias field of
initially 41G. The bias field is rotated within the plane parallel
to the Z-wire by approximately $42^{\circ}$ in order to compensate
the strong longitudinal field of the two leads of the Z-wire such
that only a small Ioffe-field remains at the trap minimum
position. The trap is compressed by increasing the bias field to
60G while forced evaporative cooling through a linear radio
frequency sweep is applied. The trap frequencies are
$\omega_{tr}=2\pi\times 150$Hz ($\omega_{tr}=2\pi\times 1.5$kHz)
and $\omega_{lo}=2\pi\times 35$Hz ($\omega_{lo}=2\pi\times 50$Hz)
for the uncompressed (compressed) trap along the transverse and
longitudinal axes, respectively. During the compression, the
transverse trap gradient is increased from 190G/cm to 450G/cm.
After 15 seconds of evaporative cooling, a Bose-Einstein
condensate of approximately $10^5$ atoms forms at a distance of
$400\mu$m from the chip surface. The details of this process are
very similar to the ones presented in \cite{Sch03,Kas03}.

To conclude, we have introduced an important simplification of
experiments with ultracold atoms near surfaces. The loading and
cooling of atoms all the way to a BEC can be achieved by
exclusively using integrated structures and small external
homogeneous bias fields. To demonstrate this, we have designed and
tested a simple Cu-structure that can be fitted underneath any
reflecting thin (up to several mm thickness are tolerable) planar
surface.

\begin{acknowledgments}
We would like to thank S. Schneider and H. Gimpel for help in the
experiments and T. Schumm for contributing to initial calculations
and designs. The atom chip used as a mirror in our experiments was
fabricated by S. Groth and I. Bar-Joseph at the Weizmann Institute
of Science (Israel). This work was supported by the European
Union, contract numbers IST-2001-38863 (ACQP) and
HPRI-CT-1999-00069 (LSF) and the Deutsche Forschungsgemeinschaft,
Schwerpunkt\-programm `Quanteninformationsverarbeitung'.
\end{acknowledgments}


\end{document}